\documentstyle[12pt]{article}
\newcommand{\thspace}{\kern.08333em}

\def \beq{\begin{equation}}
\def \eeq{\end{equation}}
\def \beqn{\begin{eqnarray}}
\def \eeqn{\end{eqnarray}}
\def \s{\sqrt{2}}

\def\to{\rightarrow}
\begin{document}
\rightline{SLAC-PUB-8221}
\rightline{hep-ph/9908343}
\rightline{August 1999}
\bigskip
\bigskip
\bigskip
\centerline {\bf CP VIOLATION AND $B$ PHYSICS \footnote{Supported 
in part by the Department of Energy under contract number DE-AC03-76SF00515.}}
\bigskip
\bigskip
\centerline {Michael Gronau~\footnote{Permanent Address: Physics 
Dept., Technion -- Israel Institute of Technology, 32000 Haifa, Israel.}}
\bigskip
\centerline{\it Stanford Linear Accelerator Center}
\centerline{\it Stanford University, Stanford, CA 94309}
\bigskip
\bigskip

\centerline{\bf ABSTRACT}
\bigskip

\begin{quote}
This is a quick review of CP non-conservation in $B$ physics. 
Several methods are described for testing the Kobayashi-Maskawa single 
phase origin of CP violation in $B$ decays, pointing out some 
limitations due to hadronic uncertainties. A few characteristic 
signatures of new physics in $B$ decay asymmetries are listed.
\end{quote}
\bigskip
\bigskip
\bigskip
\bigskip
\begin{center}
{\it Invited talk given at the 1999 Chicago Conference on Kaon Physics\\
Chicago, IL, June 21$-$26, 1999}
\end{center}

\newpage

\section{The CKM Matrix}

In the standard model of electroweak interactions CP violation is due to a 
nonzero complex phase 
\cite{KM} in the Cabibbo-Kobayashi-Maskawa (CKM) matrix $V$, describing the 
weak couplings of the charged gauge boson to quarks. 
The unitary matrix $V$, given by three mixing angles $\theta_{ij} (i<j=1,2,3)$
and a phase $\gamma$, can be approximated by ($s_{ij}\equiv \sin\theta_{ij}$)
\cite{PDG, peccei} 
\beq\label{V}
V \approx \left(\matrix{1-{1\over 2}s_{12}^2&s_{12}&s_{13}e^{-i\gamma}\cr 
-s_{12}&1-{1\over 2}s_{12}^2&s_{23}\cr 
s_{12}s_{23}-s_{13}e^{i\gamma}&-s_{23}&1\cr}\right)~. 
\eeq
Within this approximation, the only complex elements are $V_{ub}$, 
with phase $-\gamma$ and $V_{td}$, the phase of which is denoted $-\beta$.
 
The measured values of the three mixing angles and phase are \cite{peccei}
\beq\label{angles}
s_{12} = 0.220\pm 0.002~,~~s_{23} = 0.040\pm 0.003~,~~s_{13} = 0.003\pm 0.001~,
\eeq
\beq\label{gamma}
35^0\leq\gamma\equiv{\rm Arg}(V^*_{ub})\leq 145^0~.
\eeq
First evidence for a nonzero phase $\gamma$ came 35 years 
ago with the measurement of $\epsilon$, parameterizing CP violation in 
$K^0-\bar K^0$ mixing. The second evidence was obtained recently through the
measurement of ${\rm Re}(\epsilon'/\epsilon)$ \cite{eps', eps'2} discussed
extensively at this meeting.

Unitarity of $V$ implies a set of 6 triangle relations. The $db$ triangle, 
\beq\label{unit}
V_{ud}V^*_{ub} + V_{cd}V^*_{cb} + V_{td}V^*_{tb}=0~,
\eeq
is unique in having three comparable sides, which were measured
in $b\to u\ell\nu,~b\to c\ell \nu$ and $\Delta M_{d,s}$,
respectively. Whereas $V_{cb}$ was measured quite precisely, $V_{ub}$ 
and $V_{td}$ are rather poorly known at present.
The three large angles of the triangle lie in the ranges  
$35^{\circ}\le \alpha\le 120^{\circ},~10^{\circ}\le\beta\le 35^{\circ}$ and 
Eq.~(\ref{gamma}). 
As we will show in the next sections, certain $B$ decay asymmetries can 
constrain these angles considerably beyond present limits. 

For comparison with $K$ physics, note that due to the extremely small 
$t$-quark side of the $ds$ unitarity triangle 
\beq
V_{ud}V^*_{us}+V_{cd}V^*_{cs}+V_{td}V^*_{ts}=0~,
\eeq
this triangle has an angle of order $10^{-3}$, which accounts for the 
smallness of CP violation in $K$ decays. The area of this triangle,
which is equal to the area of the $db$ triangle \cite{jarl}, can be 
determined by fixing its tiny height through the rate of 
$K_L\to\pi^0\nu\bar\nu$. 
This demonstrates the complementarity of $K$ and
$B$ physics in verifying or falsifying the assumption that CP violation 
originates solely in the single phase of the CKM matrix. 

As we will show,
the advantage of $B$ decays in testing the KM hypothesis is the large variety 
of decay modes. This permits a detailed study of the phase structure of the 
CKM matrix through various interference phenomena which can measure the two 
phases $\gamma$ and $\beta$. New physics can affect this interference in 
several ways to be discussed below. 

\section{CP violation in $B^0-\bar B^0$ mixing}

The wrong-sign lepton asymmetry
\beq
A_{sl}\equiv \frac{\Gamma(\bar B^0\to X\ell^+\nu) - \Gamma(B^0\to X\ell^-
\bar\nu)}{\Gamma(\bar B^0\to X\ell^+\nu) + \Gamma(B^0\to X\ell^-\bar\nu)}~,
\eeq
measures CP violation in $B^0-\bar B^0$ mixing.
Top-quark dominance of $B^0-\bar B^0$ mixing implies that this asymmetry
is of order $10^{-3}$ or smaller \cite{bigi}.
\beq
A_{sl} = 4{\rm Re}\epsilon_B = {\rm Im}\left(\frac{\Gamma_{12}}{M_{12}}\right)
= \frac{|\Gamma_{12}|}
{|M_{12}|}{\rm Arg}\left(\frac{\Gamma_{12}}{M_{12}}\right)
\simeq \left(\frac{m^2_b}{m^2_t}\right) \left(\frac{m^2_c}{m^2_b}
\right) \leq {\cal O}(10^{-3})~.
\eeq
Present limits are at the level of 5$\%$ \cite{LEP}. 

Writing the neutral $B$ mass eigenstates as
\beq
|B_L> = p|B^0>~+~q |\bar B^0>~,~~~~|B_H> = p|B^0>~-~q |\bar B^0>~,
\eeq
one has $2{\rm Re\epsilon_B}\approx 1 - |q/p| \leq {\cal O}(10^{-3})$.
Thus, to a very high accuracy, the mixing amplitude is a pure phase
\beq
\frac{q}{p} = e^{2i{\rm Arg}(V_{td})} = e^{-2i\beta}~.
\eeq

\section{The asymmetry in $B^0(t)\to \psi K_S$}

When an initially produced $B^0$ state oscillates in time via the mixing 
amplitude which carries a phase $e^{-2i\beta}$,
\beq
|B^0(t)>~=~|B^0>\cos(\Delta mt/2) + |\bar B^0> ie^{-2i\beta}\sin(\Delta 
mt/2)~,
\eeq
the $B^0$ and $\bar B^0$ components decay with equal amplitudes to $\psi K_S$. 
The interference creates a time-dependent CP asymmetry between this process 
and the corresponding process starting with a $\bar B^0$ \cite{sanda}
\beq
A(t)=\frac {\Gamma(B^0(t)\to \psi K_S)-\Gamma(\bar B^0(t)\to \psi K_S)}
{\Gamma(B^0(t)\to \psi K_S)+\Gamma(\bar B^0(t)\to \psi K_S)}
= -\sin(2\beta)\sin(\Delta mt)~.
\eeq
The simplicity of this result, relating a measured asymmetry to an angle 
of the unitarity triangle, follows from having a single weak phase in the decay
amplitude which is dominated by $b\to c\bar c s$. This single phase
approximation holds to better than 
1$\%$ \cite{MG} and provides a clean measurement of $\beta$. 

A recent measurement by the CDF collaboration at the Tevatron \cite{kroll}, 
$\sin(2\beta) = 0.79\pm 0.39\pm 0.16$, has not yet produced a significant
nonzero result.
It is already encouraging however to note that this result prefers positive 
values, and is not in conflict with present limits, $0.4\leq\sin 2\beta\leq 
0.8$.

\section{Penguin pollution in $B^0\to\pi^+\pi^-$}

By applying the above argument to $B^0\to \pi^+\pi^-$, in which the decay 
amplitude has the phase $\gamma$, one would expect the asymmetry in this 
process to measure $\sin 2(\beta+\gamma)=-\sin(2\alpha)$. 
However, this process involves a second amplitude due
to penguin operators which carry a different weak
phase than the dominant current-current (tree) amplitude \cite{MG, LP}. 
This leads to a more general form of the time-dependent asymmetry,
which includes a new term due to direct CP violation in the decay \cite{MG}
\beq\label{asym}
A(t) = a_{\rm dir}\cos(\Delta mt) + \sqrt{1-a^2_{\rm dir}} 
\sin 2(\alpha + \theta)\sin(\Delta mt)~.
\eeq
Both $a_{\rm dir}$ and $\theta$, the correction to $\alpha$ in the second term,
are given roughly by the ratio of penguin to tree amplitudes,
$a_{\rm dir}\sim 2({\rm Penguin}/{\rm Tree})\sin\delta,~
\theta\sim ({\rm Penguin}/{\rm Tree})\cos\delta$, where $\delta$ is an 
unknown 
strong phase. A crude estimate of the penguin-to-tree ratio, based on CKM and 
QCD factors, is 0.1. Recently, flavor SU(3) was applied \cite{DGR} to relate 
$B\to \pi\pi$ to $B\to K\pi$ data, finding this ratio 
to be in the range 0.3$ \pm $0.1. Precise knowledge of this ratio could
provide very useful information about $\alpha$ \cite{MG, P/T}.

One way of eliminating the penguin effect is by measuring also the 
time-integrated rates of $B^0\to\pi^0\pi^0$, $B^+\to\pi^+\pi^0$
and their charge-conjugates \cite{GRLO}. The three $B\to\pi\pi$ amplitudes 
obey an isospin triangle relation,
\beq\label{iso}
A(B^0\to\pi^+\pi^-)/\s + A(B^0\to\pi^0\pi^0) = A(B^+\to \pi^+\pi^0)~.
\eeq
A similar relation holds for the charge-conjugate processes.
One uses the different isospin properties 
of the penguin ($\Delta I=1/2$) and tree ($\Delta I=1/2, 3/2$) contributions 
and the well-defined weak phase ($\gamma$) of the tree amplitude. This enables 
one to determine the correction to $\sin2\alpha$ in the second term of 
Eq.(\ref{asym}) by constructing the two isospin triangles.

Electroweak penguin contributions could spoil this method 
\cite{DH} since they involve $\Delta I=3/2$ components. 
This implies that the amplitudes of $B^+\to\pi^+\pi^0$ and its charge-conjugate
differ in phase, which introduces a correction at the level of a few 
percent in the isospin analysis. It was shown recently \cite{GPY} that  
this small correction can be taken into account analytically in the isospin 
analysis, since the dominant electroweak contributions are related by 
isospin to the tree amplitude. Other very small corrections can come from
isospin breaking in strong interactions \cite{gardner}.

The major difficulty of measuring $\alpha$ without knowing the ratio
Penguin/Tree is experimental rather than theoretical.
The first signal for $B^0\to\pi^+\pi^-$  reported
this summer \cite{poling, pipi}, ${\rm BR}(B^0\to\pi^+\pi^-)=
[0.47^{+0.18}_{-0.15}\pm 0.06)\times 10^{-5}$, is somewhat 
weaker than expected. Worse than that, the branching ratio
into two neutral pions is expected to be at most an order of magnitude 
smaller. This estimate is based on color-suppression, a feature already 
observed in CKM-favored $B\to \bar D\pi$ decays. Here it was found that
\cite{PDG}, ${\rm BR}(B^0\to \bar D^0\pi^0)/{\rm BR}(B^0\to D^-\pi^+)
< 0.04$. If the same color-suppression holds in $B\to\pi\pi$, then 
${\rm BR}(B^0\to \pi^0\pi^0)< 3\times 10^{-7}$, which would be too small to 
be measured with a useful precision. Constructive interference between
a color-suppressed current-current amplitude and a penguin amplitude can
increase the $\pi^0\pi^0$ rate somewhat. Limits on this rather rare mode can
be used to bound the uncertainty in determining $\sin(2\alpha)$ from 
$B^0\to\pi^+\pi^-$ \cite{GQ}
\beq
\sin(\delta\alpha)\le
\sqrt{\frac{{\cal B}(B\to \pi^0\pi^0)}{{\cal B}(B^\pm\to \pi^{\pm}\pi^0)}}
\eeq
Other ways of treating the penguin problem were discussed in \cite{other}. 

\section{$B$ decays to three pions} 

The angle $\alpha$ can also be studied in the processes $B\to\pi \rho$
\cite{rhopi}, which 
have already been seen with branching ratios larger than those of 
$B\to\pi\pi$ \cite{CLEOrhopi}, ${\rm BR}(B^0\to \pi^{\pm}\rho^{\mp})=
(3.5^{+1.1}_{-1.0}\pm 0.5)\times 10^{-5}$,~${\rm BR}(B^\pm\to \pi^{\pm}\rho^0)=
(1.5\pm 0.5 \pm 0.4)\times 10^{-5}$. 
An effective study of $\alpha$, which can eliminate
uncertainties due to penguin corrections, requires
\begin{itemize}
\item A separation between $B^0$ and $\bar B^0$ decays.
\item Time-dependent rate asymmetry measurements in $B\to\pi^{\pm}\rho^{\mp}$.
\item Measuring the rates of processes involving neutral pions, including the
color-suppressed $B^0\to \pi^0\rho^0$.
\end{itemize}
This will not be an easy task.

\section{$\gamma$ from $B \to K\pi$ and other processes}

While discussing $B^\pm$ decays to three charged pions, we note that these 
decays are of high interest for a different reason \cite{EGM}. When two 
of the pions form a mass around the charmonium $\chi_{c0}(3415)$ state, a
very large CP asymmetry is expected between $B^+$ and $B^-$ decays.
In this case the direct decay amplitude into three pions ($b\to u u\bar d$) 
interferes with a comparable amplitude into $\chi_{c0}\pi^\pm$ 
($b\to c\bar c d$) followed by $\chi_{c0}\to \pi^+\pi^-$. 
The large asymmetry (proportional to $\sin\gamma$), of order several 
tens of percent, follows from the 
$90^{\circ}$ strong phase obtained when the two pion invariant mass approaches
the charmonium mass.

A method for determining the angle $\gamma$
through $B^{\pm}\to D K^{\pm}$ decays \cite{GW}, which in principle is 
completely free of hadronic uncertainties, faces severe experimental 
difficulties. It requires measuring separately decays to states involving  
$D^0$ and $\bar D^0$. Tagging the flavor of a neutral $D$ by the charge of the
decay lepton 
suffers from a very large background from $B$ decay leptons, while tagging
by hadronic modes involves interference with doubly Cabibbo-suppressed 
$D$ decays. A few variants of this method were suggested \cite{ADS}, however,
due to low statistics, it seems unlikely that these variants can be performed 
effectively in near future facilities.

Much attention was drawn recently to studies of $\gamma$ in 
$B\to K\pi$, motivated by measurements of charge-averaged $B\to K\pi$ decay 
branching ratios \cite{poling, pipi}
\beqn
{\rm BR}(B^\pm\to K\pi^\pm) & = & (1.82^{+0.46}_{-0.40}\pm 0.16)
\times 10^{-5}~,\\
\nonumber
{\rm BR}(B^\pm\to K^\pm\pi^0) & = & (1.21^{+0.30+0.21}_{-0.28-0.14})
\times 10^{-5}~,\\
\nonumber
{\rm BR}(B^0\to K^\pm\pi^\mp) & = & (1.88^{+0.28}_{-0.26}\pm 0.13)
\times 10^{-5}~,\\
\nonumber
{\rm BR}(B^0\to K^0\pi^0) & = & (1.48^{+0.59+0.24}_{-0.51-0.33})
\times 10^{-5}~.
\eeqn
The first suggestion to constrain $\gamma$ from $B\to K\pi$ was made
in \cite{GRL}, where electroweak penguin contributions were neglected.
The importance of electroweak penguin terms was noted in \cite{DHF}, which 
was followed by several ideas about controlling these effects \cite{EWP}.
In the present discussion we will focus briefly on very recent work
along these lines \cite{GPY, NR, GP, BF}, simplifying the discussion
as much as possible. 

Decomposing the $B^+\to K\pi$ amplitudes into contributions
from penguin ($P$), color-favored tree ($T$) and color-suppressed tree 
($C$) terms \cite{GHLR},
\beq\label{PTC}
A(B^+\to K^0\pi^+)=P~,~~~A(B^+\to K^+\pi^0)=-(P+T+C)/\s~,
\eeq
$P$ has a weak phase $\pi$, while $T$ and $C$ each carry the phase 
$\gamma$.
Some information about the relative magnitudes of these terms can be gained
by using SU(3) and comparing these amplitudes to those of $B\to \pi\pi$ 
\cite{DGR}. This implies
\beq
r\equiv \frac{T+C}{P} = 0.24\pm 0.06~.
\eeq
Defining the ratio of charge-averaged rates \cite{NR}
\beq
R^{-1}_*=\frac{2{\cal B}(B^{\pm}\to K^{\pm}\pi^0)}
{{\cal B}(B^{\pm}\to K\pi^{\pm})}~,
\eeq
one has
\beq\label{noEWP}
R^{-1}_* = 1 - 2r\cos\delta\cos\gamma + r^2~,
\eeq
where $\delta$ is the penguin-tree strong phase-difference.
Any deviation of this ratio from one would be a clear signal of interference 
between $T+C$ and $P$ in $B^+\to K^+\pi^0$ and could be used to constrain 
$\gamma$.

So far, electroweak penguin contributions have been neglected. These terms 
can be 
included in the above ratio by relating them through flavor SU(3) to the 
corresponding tree amplitudes. This is possible since the two types
of operators have the same (V-A)(V-A) structure and differ only by SU(3). 
Hence, in the SU(3) limit, the dominant electroweak penguin term
and the tree amplitude have the same strong phase, and the ratio of 
their magnitudes is given simply by a ratio of the corresponding Wilson
coefficients multiplied by CKM factors \cite{GPY, NR}
\beqn\label{del}
\delta_{EW} & \equiv & \frac{|{\rm EWP}(B^+\to K^0\pi^+) + \sqrt{2}{\rm EWP}
(B^+\to K^+\pi^0)|}{|T+C|}\\
& = & -\frac{3}{2}\frac{c_9 + c_{10}}{c_1 + c_2}\frac{|V^*_{tb}V_{ts}|}
{|V^*_{ub}V_{us}|} = 0.6 \pm 0.2~,
\eeqn
where the error comes from $|V_{ub}|$.
Consequently, one finds instead of (\ref{noEWP})
\beq
R^{-1}_* = 1 - 2r \cos\delta (\cos\gamma - \delta_{EW}) + {\cal O}(r^2)~,
\eeq
implying
\beq
|\cos\gamma - \delta_{EW}| \ge \frac{|1-R^{-1}_*|}{2r}
\eeq
{\it If} $R^{-1}_*\ne 1$, this constraint can be used to exclude a region 
around $\gamma=50^{\circ}$. The present value of $R^{-1}_*$ is consistent
with one. Experimental errors must be substantially reduced before drawing
any conclusions.

The above constraint is based only on charge-averaged rates.
Further information on $\gamma$ can be obtained by measuring separately $B^+$
and $B^-$ decay rates. The $B^+ \to K\pi$ rates obey a triangle relation
with $B^+\to\pi^+\pi^0$ \cite{GPY, GRL, NR}
\beq
\sqrt2 A(B^+\to K^+\pi^0) + A(B^+\to K^0\pi^+) =
\tilde r_u A(B^+\to\pi^+\pi^0) \left(1 - \delta_{EW} e^{-i\gamma}\right)~,
\eeq
where $\tilde r_u = (f_K/f_{\pi})\tan\theta_c\simeq 0.28$ contains explicit
SU(3) breaking. This relation and its charge-conjugate permit a determination 
of $\gamma$ which does not rely on $R^{-1}_*\ne 1$.

This analysis involves uncertainties due to errors in $r$ and 
$\delta_{EW}$, which are expected to be reduced to the level of 10$\%$. 
Additional uncertainties follow from SU(3) breaking in (\ref{del}) and
from rescattering effects in $B^+\to K^0\pi^+$ which introduce a term with
phase $\gamma$ in this process. The latter effects can be bounded by the
U-spin related rate of $B^+\to K^+ \bar K^0$ \cite{rescat}. Present limits on 
rescattering corrections are at a level of 20$\%$ and can be reduced to 
10$\%$ in future high statistics experiments. Such rescattering corrections
introduce an error of about $10^{\circ}$ in determining $\gamma$ 
\cite{GP}. Summing up all the theoretical uncertainties, and neglecting 
experimental errors, it is unlikely that this method will determine $\gamma$ 
to better than $\pm 20^{\circ}$. Nevertheless, this 
would be a substantial improvement relative to the present bounds 
(\ref{gamma}).

We conclude this section with a simple observation \cite{MJR}, which 
enables an early detection of a CP asymmetry in $B\to K\pi$.
Using $A(B^0\to K^+\pi^-)=-P-T$, the hierarchy among amplitudes \cite{GHLR}, 
$|P|\gg |T| \gg |C|$, implies ${\rm Asym}(B^{\pm}\to K^{\pm}\pi^0) \approx 
{\rm Asym}(B\to K^{\pm}\pi^{\mp})$. This may be used to 
gain statistics by measuring the combined asymmetry in these two modes.
The magnitude of the asymmetry depends on an unknown final state strong phase.
Very recently a 90$\%$ confidence level upper limit was reported
${\rm Asym}(B\to K^{\pm}\pi^{\mp}) < 0.35$ \cite{poling, asym}.

\section{Signals of new physics}

The purpose of future $B$ physics is to over-constrain the unitarity triangle.
$|V_{ub}|$ can at best be determined to 10$\%$ \cite{liget} and $|V_{td}|$
relies on future measurements of the higher order $B^0_s-\bar B^0_s$ 
mixing \cite{kroll} and $K^+\to \pi^+\nu\bar\nu$ \cite{red}. Constraining 
the angles $\alpha,~\beta$ 
and $\gamma$ by CP asymmetries is complementary to these CP conserving 
measurements. The asymmetry measurements involve discrete ambiguities in the 
angles, which ought to be resolved \cite{wolf}.

Hopefully, these studies will not only sharpen our knowledge of the CKM
parameters but will eventually show some inconsistencies. In this case, 
the first purpose of $B$ physics will be to identify the source 
of the inconsistencies in a model-independent way. Let us discuss 
this scenario briefly by considering a few general possibilities.
 
Physics beyond the standard model can modify CKM phenomenology and predictions
for CP asymmetries by introducing additional contributions in three types of 
amplitudes:
\begin{itemize}
\item $B^0-\bar B^0$ and $B^0_s-\bar B^0_s$ mixing amplitudes.
\item Penguin decay amplitudes.
\item Tree decay amplitudes.
\end{itemize}
The first case is the most likely possibility, demonstrated by a large 
variety of models \cite{nir}. New mixing terms, which can be large and which
often also affect the rates of electroweak penguin decays,
modify in a universal way the interpretation of asymmetries in terms of
phases of $B^0-\bar B^0$ and $B^0_s-\bar B^0_s$ mixing amplitudes. These 
contributions can be identified either 
by measuring asymmetries which lie outside the allowed range, 
or by comparison with mixing-unrelated constraints. On the other hand, new 
contributions in decay amplitudes \cite{gw} are usually small, may 
vary from one process to another, and can be detected be comparing 
asymmetries in different processes. Processes in which the KM hypothesis 
implies extremely small asymmetries are particularly sensitive to new
amplitudes. 

To conclude this brief discussion, let us list a few examples of signals for 
new physics.
\begin{itemize}
\item $A_{sl} \geq {\cal O}(10^{-2})$.
\item Sizable asymmetries in $b \to s \gamma$ or $B_s\to \psi\phi$.
\item ``Forbidden'' values of angles, $|\sin 2\beta - 0.6| > 0.2,~~
\sin\gamma < 0.6$.
\item Different asymmetries in $B^0(t)\to \psi K_S,~\phi K_S,~\eta' K_S$.
\item Contradictory constraints on $\gamma$ from $B\to K\pi,~
B\to D K,~B_s\to D_sK$.
\item Rate enhancement beyond standard model predictions for electroweak 
penguin decays, $B\to X_{d,s}\ell^+\ell^-,~B^0/B_s\to\ell^+\ell^-$.
\end{itemize}

\section{Conclusion}

The CP asymmetry in $B\to \psi K_S$ is related cleanly 
to the weak phase $\beta$ and can be used experimentally to measure 
$\sin 2\beta$. In other cases, such as in $B^0\to\pi^+\pi^-$ which 
measures $\sin 2\alpha$ and $B \to DK$ which determines $\sin\gamma$,
the relations between the asymmetries, supplemented by certain rates, and the 
corresponding weak phases are
free of significant theoretical uncertainties. However, the application of
these methods are expected to suffer from experimental difficulties due
to the small rates of color-suppressed processes. 

While one expects qualitatively that color-supression
is affected by final-state interactions, these long distance phenomena 
are not understood quantitatively.
The case of $B\to K\pi$ demonstrates the need for a better undersanding of 
these features, and the need for a reliable treatment 
of SU(3) breaking. That is, whereas the short distance effects of QCD in weak 
hadronic $B$ decays are well-understood \cite{BBL}, we are in great need 
of a theoretical
framework for studying long distance effects. An interesting suggestion in
this direction was made very recently in \cite{beneke}.

We discussed mainly the very immediate $B$ decay 
modes, for which CP asymmetries can provide new information on CKM parameters.
Asymmetries should be searched in {\it all $B$ decay
processes}, including those which are plagued by theoretical uncertainties 
due to unknown final state interactions, and those where the KM framework 
predicts negligibly small asymmetries. Afterall, our understanding of the
origin of CP violation is rather limited and surprises may be right around 
the corner.  

\bigskip
{\it Acknowledgments}: I thank the SLAC Theory Group for its very kind 
hospitality. I am grateful to Gad Eilam, David London, Dan Pirjol, 
Jon Rosner and Daniel Wyler for collaborations on topic discussed here.
This work was supported in part 
by the United States $-$ Israel Binational Science Foundation under research
grant agreement 94-00253/3, and by the Department of Energy under contract
number DE-AC03-76SF00515.

\newpage
\def \ijmp#1#2#3{{\it Int.~J.~Mod.~Phys.}~A {\bf#1}, #2 (#3)}
\def \np#1#2#3{{\it Nucl.~Phys.}~B {\bf#1}, #2 (#3)}
\def \plb#1#2#3{{\it Phys.~Lett.}~B {\bf#1}, #2 (#3)}
\def \prd#1#2#3{{\it Phys.~Rev.}~D {\bf#1}, #2 (#3)}
\def \prl#1#2#3{{\it Phys.~Rev.~Lett.}~{\bf#1}, #2 (#3)}
\def \prp#1#2#3{{\it Phys.~Rep.}~{\bf#1} #2 (#3)}
\def \ptp#1#2#3{{\it Prog.~Theor.~Phys.}~{\bf#1}, #2 (#3)}
\def \rmp#1#2#3{{\it Rev.~Mod.~Phys.}~{\bf#1} #2 (#3)}
\def \zpc#1#2#3{{\it Z.~Phys.}~C {\bf#1}, #2  (#3)}
\def \ite{{\it et al.}}
\def \stone{{\it B Decays}, (World Scientific, 1994)}
\def \epj#1#2#3{{\it Eur. Phys. J.}~C {\bf#1}, #2 (#3)}


\begin{thebibliography}{99}
\bibitem{KM} M. Kobayashi and T. Maskawa, \ptp{49}{652}{1973}.
\bibitem{PDG} C. Caso \ite, \epj{3}{1}{1998}.
\bibitem{peccei} R. Peccei, these proceedings, also discusses the
Wolfenstein parameterization in terms of $\lambda,~A,~\rho$ and
$\eta$.
\bibitem{eps'} Y. B. Hsiung, these proceedings.
\bibitem{eps'2} M. Sozzi, these proceedings.  
\bibitem{jarl} C. Jarlskog, \prl{55}{1039}{1985}. 
\bibitem{bigi} I. I. Bigi \ite, in {\it CP Violation}, ed. C. Jarlskog 
(World Scientific, Singapore, 1992).
\bibitem{LEP} OPAL Collaboration, K. Ackerstaff \ite, \zpc{76}{401}{1997}.
\bibitem{sanda} A. B. Carter and A. I. Sanda, \prl{45}{952}{1980}; \prd{23}
{1567}{1981}; I. I. Bigi and A. I. Sanda, \np
{193}{85}{1981}. 
\bibitem{MG} M. Gronau, \prl{63}{1451}{1989}.
\bibitem{kroll} J. Kroll, these proceedings.
\bibitem{LP}D. London and R. D. Peccei, \plb{223}{257}{1989}; B. Grinstein, 
\plb{229}{280}{1989}.
\bibitem{DGR} A. Dighe, M. Gronau and J. L. Rosner, \prl{79}{4333}{1997}.
\bibitem{P/T} F. DeJongh and P. Sphicas, \prd{53}
{4930}{1996}; P. S. Marrocchesi and N. Paver, \ijmp{13}{251}{1998}. 
\bibitem{GRLO} M. Gronau and D. London, \prl{65}{3381}{1990}.
\bibitem{DH} N. G. Deshpande and X. G. He, {\it Phys. Rev. Lett.}
{\bf 74}, 26, 4099(E) (1995).
\bibitem{GPY} M. Gronau D. Pirjol and T. M. Yan, \prd{60}{034021}{1999}.
\bibitem{gardner} S. Gardner, \prd{59}{077502}{1999}.
\bibitem{poling} R. Poling, Rapporteur talk at the 19th International
Lepton-Photon Symposium, Stanford, CA, August 9$-$14, 1999.
\bibitem{pipi} CLEO Collaboration, Y. Kwon \ite, hep-ex/9908029. 
\bibitem{GQ} Y. Grossman and H. R. Quinn, \prd{58}{017504}{1998}.
\bibitem{other} J. Charles, \prd{59}{054007}{1999}; D. Pirjol, 
\prd{60}{54020}{1999}; R. Fleischer, hep-ph/9903456.
\bibitem{rhopi} H. J. Lipkin, Y. Nir, H. R. Quinn and A. Snyder, 
\prd{44}{1454}{1991}; M. Gronau, \plb{265}{389}{1991}; H. R. Quinn and A. 
Snyder, \prd{48}{2139}{1993}.
\bibitem{CLEOrhopi} CLEO Collaboration, M. Bishai \ite, hep-ex/9908018.
\bibitem{EGM} G. Eilam, M. Gronau and R. R. Mendel, \prl{74}{4984}{1995};
N. G. Deshpande \ite, \prd{52}{5354}{1995}; I. Bediaga, R. E. Blanco, C. 
Gobel and R. Mendez-Galain, \prl{81}{4067}{1998}; B. Bajc \ite, \plb{447}
{313}{1999}.
\bibitem{GW} M. Gronau and D. Wyler, \plb{265}{172}{1991}.
\bibitem{ADS} D. Atwood, I. Dunietz and A. Soni, \prl{78}{3257}{1997};
M. Gronau, \prd{58}{037301}{1998}; M. Gronau and J. L. Rosner, \plb{439}{171}
{1998}. 
\bibitem{GRL} M. Gronau, J. L. Rosner and D. London, \prl{73}{21}{1994}.
\bibitem{DHF} N. G. Deshpande and X. G. He, \prl{74}{26}{1995}.
Electroweak penguin effects in other $B$ decays were studied earlier by
R. Fleischer, \plb{321}{259}{1994}.
\bibitem{EWP} R. Fleischer, \plb{365}{399}{1996}; A. J. 
Buras and R. Fleischer, \plb{365}{390}{1996}; M. Gronau and J. L. Rosner, 
\prl{76}{1200}{1996}; A. S. Dighe, M. Gronau and J. L. Rosner, 
\prd{54}{3309}{1996}; R. Fleischer and T. Mannel, \prd{57}{2752}{1998}.
\bibitem{NR} M. Neubert and J. L. Rosner, \plb{441}{403}{1998}; \prl{81}{5076}
{1998}; M. Neubert, JHEP {\bf 9902}, 014 (1999). 
\bibitem{GP} M. Gronau and D. Pirjol, hep-ph/9902482.
\bibitem{BF} A. J. Buras and R. Fleischer, hep-ph/9810260.
\bibitem{GHLR} M. Gronau, O. Hern\'andez, D. London and J. L. Rosner,
\prd{50}{4529}{1994};  \prd{52}{6374}{1995}.
\bibitem{rescat} A. Falk, A. L. Kagan, Y. Nir and A. A. Petrov, \prd{57}
{4290}{1998}; M. Gronau and J. L. Rosner, \prd{57}{6843}{1998}; {\bf 58}, 
113005 (1998); R. Fleischer, \plb{435}{221}{1998}; \epj{6}{451}{1999}.
\bibitem{MJR} M. Gronau and J. L. Rosner, \prd{59}{113002}{1999}.
\bibitem{asym} CLEO Collaboration, T. E. Coan \ite, hep-ex/9908029.
\bibitem{liget} Z. Ligeti, these proceedings.
\bibitem{red} G. Redlinger, these proceedings.
\bibitem{wolf} Y. Grossman and H. R. Quinn, \prd{56}{7529}{1997};
L. Wolfenstein, \prd{57}{6857}{1998}.
\bibitem{nir} C. O. Dib, D. London and Y. Nir, \ijmp{6}{1253}{1991}; M. Gronau 
and D. London, \prd{55}{2845}{1997}.
\bibitem{gw} Y. Grossman and M. P. Worah, \plb{395}{241}{1997}; D. London and 
A. Soni, \plb{407}{61}{1997}.
\bibitem{BBL} G. Buchalla, A. J. Buras and M. E. Lautenbacher, 
\rmp{68}{1125}{1996}.
\bibitem{beneke} M. Beneke, G. Buchalla, M. Neubert and C. T. Sachrajda, 
hep-ph/9905312.

\end{thebibliography}
\end{document}